\documentstyle[12pt,aaspp4]{article}
\begin{document}

\title{The \it Ulysses \rm \bf Supplement to the BATSE 4Br Catalog of Cosmic
Gamma-Ray Bursts}
\author{K. Hurley}
\affil{University of California, Berkeley, Space Sciences Laboratory,
Berkeley, CA 94720-7450}
\authoremail{khurley@sunspot.ssl.berkeley.edu}
\author{M. S. Briggs, R. M. Kippen}
\affil{University of Alabama in Huntsville, Huntsville AL 35899}
\author{C. Kouveliotou\altaffilmark{1}, C. Meegan, G. Fishman} 
\affil{NASA Marshall Space Flight Center, Huntsville AL 35812}
\author{T. Cline}
\affil{NASA Goddard Space Flight Center, Code 661, Greenbelt, MD 20771}
\author{M. Boer}
\affil{Centre d'Etude Spatiale des Rayonnements, B.P. 4346, 31029
Toulouse, France}

\altaffiltext{1}{Universities Space Research Association, Marshall Space
Flight Center ES-84, Huntsville, AL 35812}

\begin{abstract}

We present Interplanetary Network localization information for 147 gamma-ray bursts
observed by the Burst and Transient Source Experiment between the end of 
the 3rd BATSE catalog and the end of the 4th BATSE catalog, obtained by analyzing the arrival times of
these bursts at the \it Ulysses \rm and \it Compton Gamma-Ray Observatory \rm (CGRO) spacecraft.  
For any given
burst observed by these two spacecraft, arrival time analysis
(or ``triangulation'') results in an annulus of possible arrival
directions whose half-width varies between 7 arcseconds and 2.3 degrees, depending
on the intensity and time history of the burst, and the distance of the \it Ulysses \rm
spacecraft from Earth.  This annulus generally
intersects the BATSE error circle, resulting in an average reduction of the
error box area of a factor of 25.  

\end{abstract}


\section{Introduction}

In a previous paper (Hurley et al. 1998) we have presented the \it Ulysses
\rm supplement to the BATSE 3B Catalog of Cosmic Gamma-Ray Bursts (hereafter
referred to as the 3B supplement).  This catalog
contained improved positions for 218 bursts in the 3rd BATSE catalog (Meegan et 
al. 1996), obtained by arrival time analysis, or ``triangulation'' between
the \it Compton Gamma-Ray Observatory \rm and \it Ulysses \rm spacecraft.
The last BATSE burst in the 3B supplement is \#3168, on 1994 September 15.  In
the present catalog, we update the 3B supplement with the addition of data
on 147 bursts which occurred between 1994 September 15 and 1996 August 15 (BATSE
\#5575),
the last burst in the BATSE 4Br catalog (Paciesas et al. 1998) observed by Ulysses.  
As none of the
information in the 3B supplement has changed, we do not include any of the
detailed data on the bursts in that catalog.

\section{Instrumentation, Search Procedure, Derivation of Annuli, and
Burst Selection Criteria}

None of these has changed from the 3B supplement.  We review each briefly,
but refer the reader to Hurley et al. (1998) for a more detailed description.

The \it Ulysses \rm GRB detector (Hurley et al. 1992) consists of two 3 mm thick
hemispherical CsI scintillators
with a projected area of about 20 cm$^2$ in any direction.  The detector is mounted
on a magnetometer boom far from the body of the spacecraft, and has a practically
unobstructed view of the full sky.  BATSE consists of eight detector modules situated at the corners of the \it Compton
Gamma-Ray Observatory \rm spacecraft.  Each contains a Large Area Detector (LAD), a 
50.8 cm diameter by 1.27 cm thick
NaI scintillator  (Meegan et al. 1996).  

Every cosmic burst detected by BATSE is systematically searched for in the
\it Ulysses \rm data as soon as the BATSE data are available for it.
This is done by using the approximate arrival direction
from BATSE and the position of the \it Ulysses \rm spacecraft to calculate
a range of possible arrival times at \it Ulysses \rm.  Typical window lengths are 300 - 500 s.

When a GRB arrives at two spacecraft with a delay $\rm \delta$T, it may be
localized to an annulus whose half-angle $\rm \theta$ with respect to the
vector joining the two spacecraft is given by 
\begin{equation}
cos \theta=\frac{c \delta T}{D}
\end{equation}
where c is the speed of light and D is the distance between the two
spacecraft.  (This assumes that the burst is a plane wave, i.e. that its
distance is much greater than D.)  The annulus width d$\rm \theta$, and thus one dimension of
the resulting error box, is c $\rm \sigma(\delta$T)/Dsin$\rm \theta$ where
$\rm \sigma(\delta$T) is the uncertainty in the time delay.  
The radius of each annulus and its right ascension and declination are
calculated in a heliocentric (i.e., aberration-corrected) frame.  

The main selection criterion for a burst in this catalog is that it 
must have been detected by \it Ulysses \rm and
BATSE.  We also utilize several other criteria, based on the correlation
coefficient between the two light curves, a chi-squared statistic, and the
ratio of the observed \it Ulysses \rm counts to the BATSE counts.
A final criterion for inclusion in the 3B supplement, namely that the burst must
not have been observed by a third interplanetary spacecraft, is automatically
satisfied for the present catalog, since no interplanetary spacecraft with
burst detectors were in operation, other
than Ulysses.  

\section{A Few Statistics}

There are 1637 bursts in the 4Br catalog (Paciesas et al. 1998).  Of these, 
428 were observed by \it Ulysses \rm and BATSE and
in some cases, other spacecraft as well \footnote{.  
A list of all cosmic bursts and the spacecraft which detected them may be found
at http://ssl.berkeley.edu/ipn3/index.html}.
Thus \it Ulysses \rm observed
approximately one out of every 3.8 BATSE bursts over this period.  The
combination of the 3B supplement and the present catalog contains 365 bursts. 

The histogram of Figure 1 shows the distribution of annulus half-widths 
(i.e. $\rm \delta R_{IPN}$ in Table 1) for the
365 bursts localized.  The smallest is about 7 \arcsec, the
largest 2.3$^{\rm o}$, and the average is 5.4 \arcmin.  265 of the annuli,
or 73\%, intersect the BATSE 1 $\sigma$ error circles, whose radii are defined by 
$\rm r_{1\sigma}=\sqrt{\sigma_{stat}^2 + \sigma_{sys}^2}$,
where $\sigma_{sys}$ is the systematic error, 1.6$^{\rm o}$,
and $\sigma_{stat}$ is the statistical error.  This is less than
the number which would be predicted (87\%).  An analysis of a preliminary
version of the IPN catalog describes several more complicated BATSE error
models that are consistent with the BATSE-IPN separations (Briggs et al. 1998a).
A more extensive analysis, utilizing the data in the present paper, 
appears in Briggs et al. 1998b.
One quantity of interest is how close the
annulus passes to the center of the error circle.  Let 
$\alpha_1, \delta_1$ be the right ascension and declination of
the center of a BATSE error circle, and let $\alpha_2, \delta_2, \theta_2$
be the right ascension, declination, and radius of an annulus.
Then the minimum distance between the error circle and the annulus
is given by
\begin{equation}
d=\mid \theta_2 - \cos^{-1}(\sin(\delta_1) \sin(\delta_2) +
\cos(\delta_1) \cos(\delta_2) \sin(\alpha_1 - \alpha_2) ) \mid 
\end{equation}
A histogram of the distribution of minimum distances
between the annuli and the centers of the BATSE error circles
is given in Figure 2.

In general, the annuli obtained by triangulations are small circles on the celestial
sphere, so their curvature, even across a relatively small BATSE error circle, is
not always negligible, and a simple, four-sided error box cannot be defined. 
For this reason, we do not cite the intersection points of
the annulus with the error circle.  

Figure 3 shows the BATSE peak fluxes and fluences for 285 of the 365 bursts
with flux and fluence entries in the 4Br catalog.  

\section{Table of Annuli}

The ten columns in table 1 give:
1) the date of the burst, in ddmmyy format, 
2) the Universal Time of the burst at Earth,
3) the BATSE number for the burst,
4) the BATSE right ascension of the center of the error circle (J2000), in degrees,
5) the BATSE declination of the center of the error circle (J2000), in degrees,
6) the total 1 $\sigma$ statistical BATSE error circle radius, in degrees, (the 
approximate total
1$\sigma$ radius is obtained by adding 1.6$^{\rm o}$ in quadrature, but see
Briggs et al. 1998a,b for an improved error model), 
7) the right ascension of the center of the IPN (BATSE/\it Ulysses \rm)
annulus, epoch J2000, in the heliocentric frame, in degrees,
8) the declination of the center of the IPN (BATSE/\it Ulysses \rm)
annulus, epoch J2000, in the heliocentric frame, in degrees, 
9) the angular radius of the IPN (BATSE/\it Ulysses \rm) annulus, in the heliocentric
frame, in degrees, and
10) the half-width of the IPN (BATSE/\it Ulysses \rm) annulus, in degrees; the 3 $\sigma$
confidence annulus is given by R$_{\rm IPN}$ $\pm$ $\delta$ R$_{IPN}$.

Entries are given only for the 147 bursts which occurred between the end
of the 3B and the end of the 4Br catalog.
The BATSE data have been taken from the latest online catalog,
and are given here for convenience only;
the catalog \footnote{available at http://www.batse.msfc.nasa.gov/data/grb/4bcatalog/}
should be considered to be the ultimate source of the
most up-to-date BATSE data.   Table 1 is also available electronically \footnote
{at http://ssl.berkeley.edu/ipn3/index.html}.  

Figure 4 compares the BATSE error circles with the IPN annulus/error circle
intersections for the bursts in this catalog.  To generate the plot, it was 
assumed that all annuli pass through the centers of their corresponding BATSE error 
circles.

\section{Conclusion}

The \it Ulysses \rm GRB experiment continues to operate.  As of 1998 April, it
has detected about 560 BATSE gamma-ray bursts.  Data on these events may be found at
the IPN web site \footnote{http://ssl.berkeley.edu/ipn3/index.html}.

\section{Acknowledgments}

Support for the \it Ulysses \rm GRB experiment is provided by JPL Contract 958056.  Joint
analysis of \it Ulysses \rm and BATSE data is supported by NASA Grant NAG 5-1560.  CK acknowledges
support from NASA Grant NAG5-2560.

\clearpage

\begin{figure}
\plotone{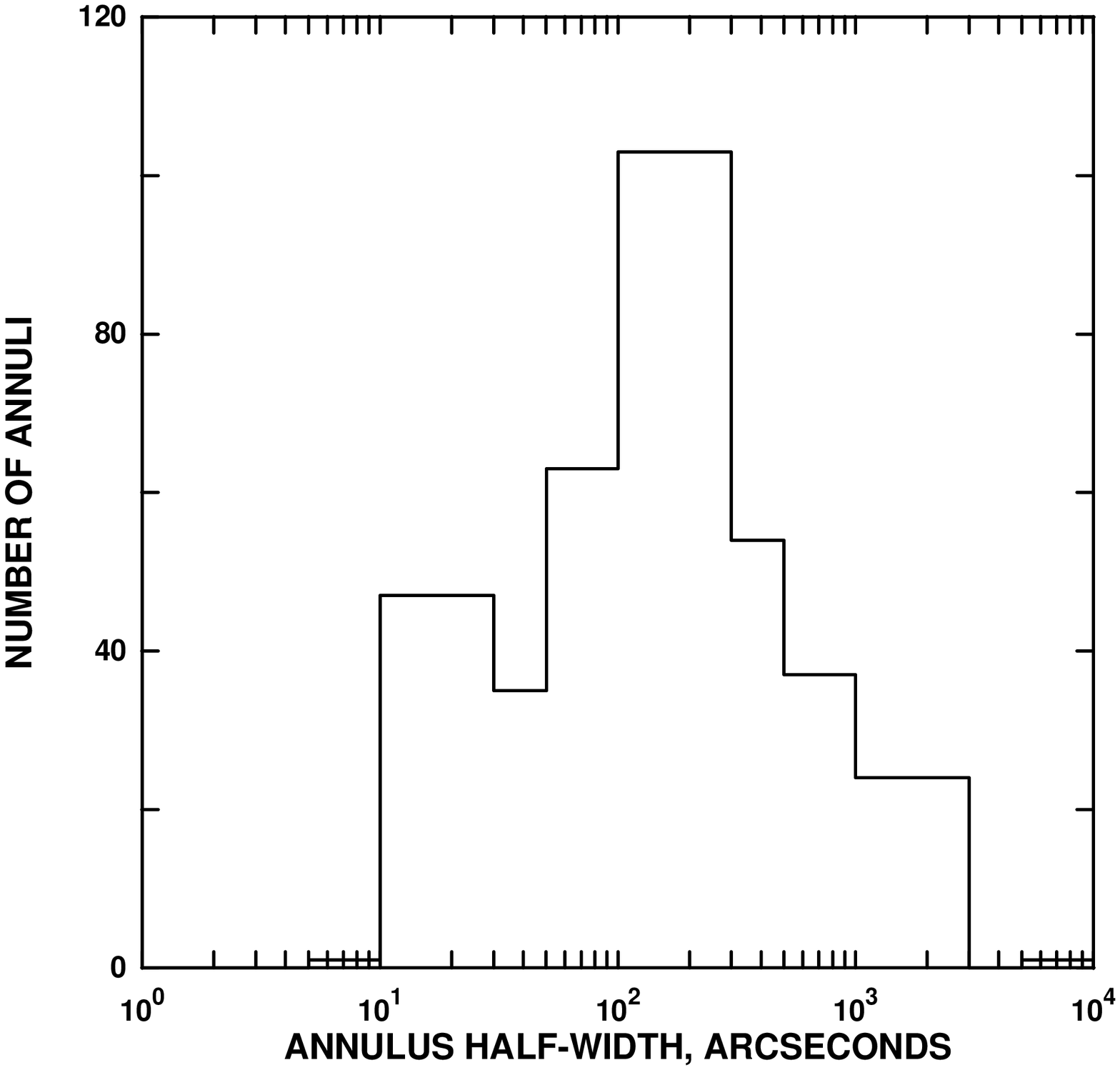}
\caption{Distribution of 365 \it Ulysses \rm/BATSE annulus half-widths. \label{Fig. 1}}
\end{figure}

\begin{figure}
\plotone{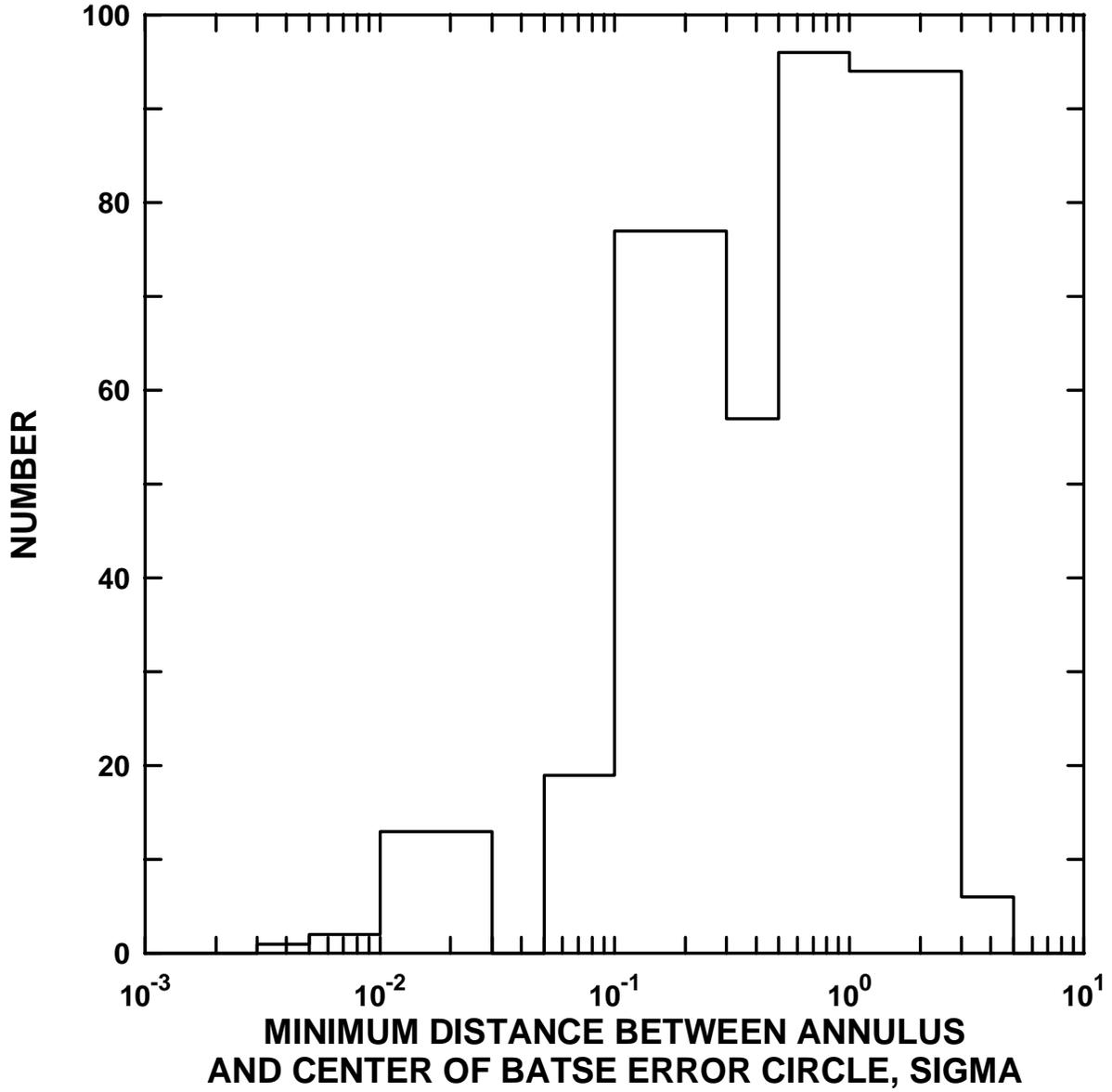}
\caption{Distribution of minimum angular distances between the 365 annuli
and the centers of the BATSE error circles.  The angular distances are expressed in
number of sigma for the BATSE error circle radii, i.e.
d/r$_{1\sigma}$. \label{Fig. 2}}
\end{figure}

\begin{figure}
\plotone{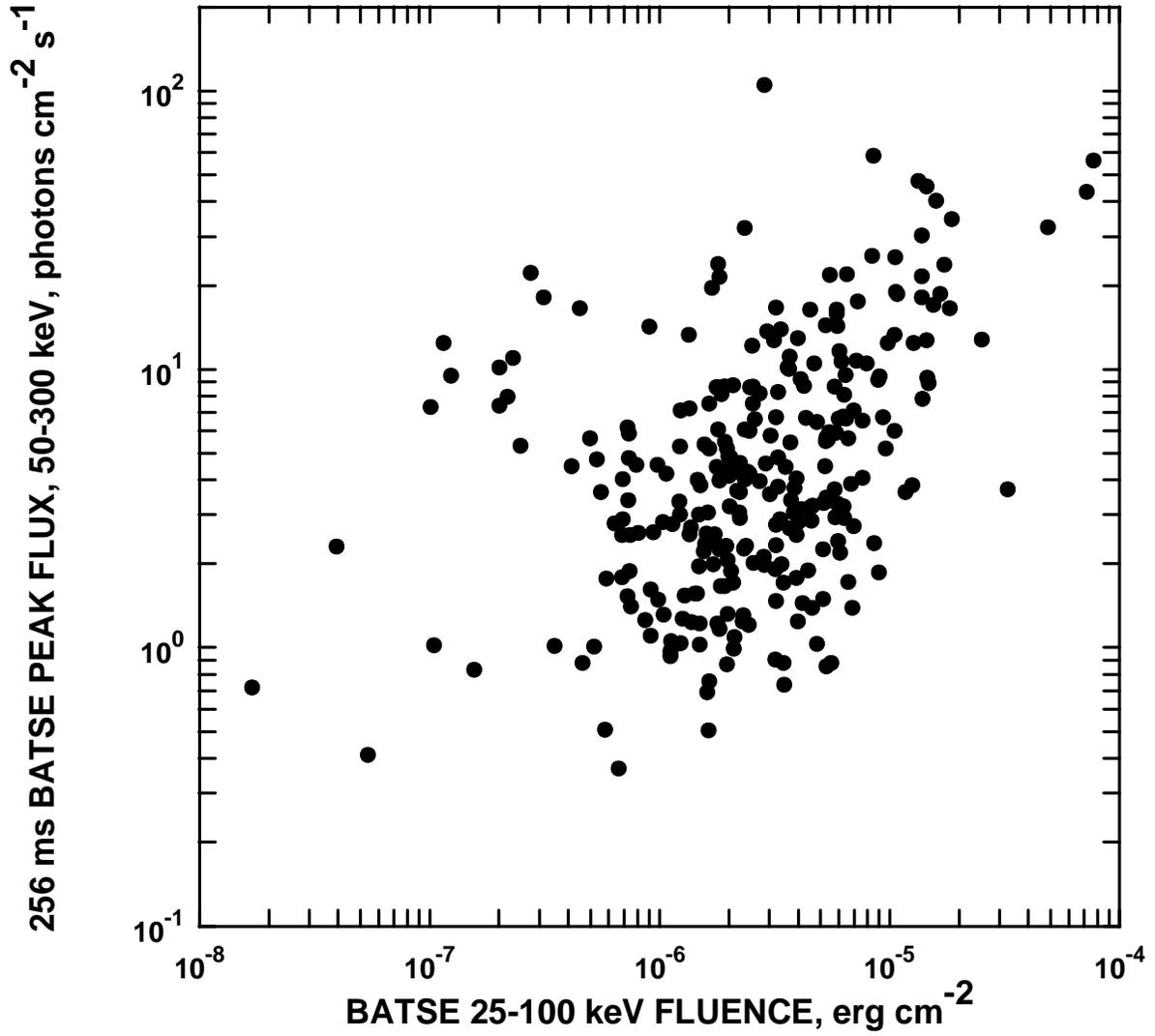}
\caption{Peak fluxes (measured over 256 ms, 50-300 keV) and 25-100 keV fluences of
285 of the bursts in this catalog.  No entries are given in the 4Br catalog for 80
of the bursts. \label{Fig. 3}}
\end{figure}

\begin{figure}
\plotone{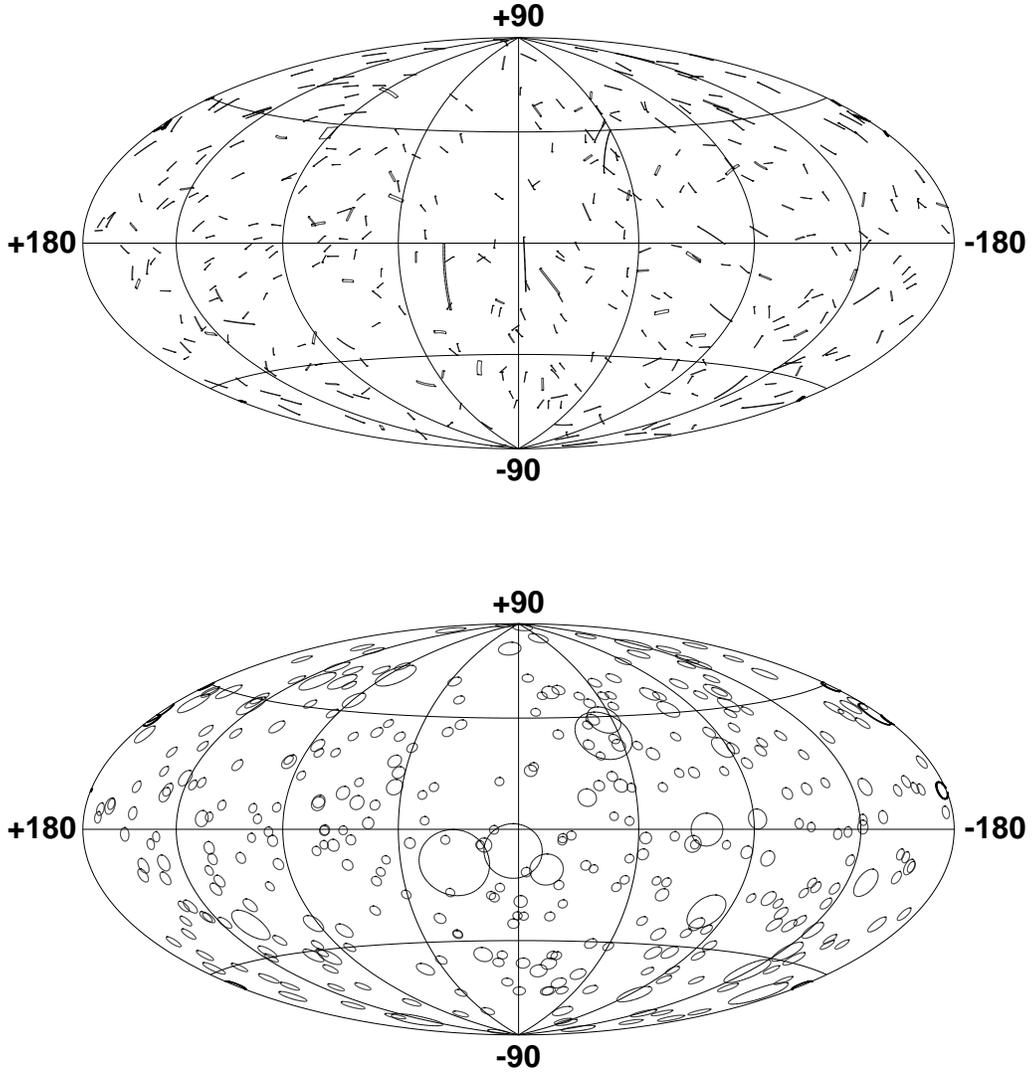}
\caption{Top:  Distribution of 365 \it Ulysses \rm/BATSE error circle/annulus intersections.
bottom: distribution of 365 BATSE 4Br error circles in galactic coordinates.
\label{Fig. 4}}
\end{figure}

\clearpage
\begin{deluxetable}{cccccccccc}
\tablecaption{\it Ulysses \rm/BATSE annuli}
\tablehead{
\colhead{Date}&\colhead{UT}&\colhead{N$_B$}&\colhead{$\alpha_{2000, B}$}&
\colhead{$\delta_{2000, B}$}&\colhead{$\sigma_{stat, B}$}&\colhead{$\alpha_{2000, IPN}$}&
\colhead{$\delta_{2000, IPN}$}&\colhead{R$_{IPN}$}&\colhead{$\delta R_{IPN}$}
}
\startdata
210994 & 05:08:13 & 3178 & 204.90 & -17.77 & 5.29 & 155.521 & -55.212 & 54.343 & 0.014 \nl
031094 & 05:17:17 & 3212 & 186.90 & 35.04 & 1.29 & 343.275 & 60.048 & 79.473 & 0.094 \nl
081094 & 13:32:52 & 3227 & 305.62 & -52.18 & 0.87 & 167.522 & -62.363 & 58.853 & 0.011 \nl
141094 & 09:20:08 & 3241 & 31.66 & 7.43 & 1.14 & 352.973 & 64.966 & 64.035 & 0.007 \nl
141094 & 23:15:16 & 3242 & 164.78 & 19.74 & 1.31 & 173.571 & -65.225 & 84.182 & 0.027 \nl
171094 & 10:19:33 & 3245 & 301.55 & 9.81 & 0.19 & 356.264 & 66.336 & 65.561 & 0.013 \nl
201094 & 18:20:00 & 3251 & 158.99 & -2.72 & 2.33 & 180.319 & -67.826 & 69.752 & 0.157 \nl
201094 & 20:05:27 & 3253 & 49.99 & -24.74 & 1.03 & 180.416 & -67.860 & 82.022 & 0.015 \nl
231094 & 04:34:32 & 3255 & 212.28 & -82.35 & 0.42 & 183.608 & -68.895 & 15.870 & 0.031 \nl
261094 & 02:52:35 & 3257 & 205.96 & -6.53 & 0.47 & 188.047 & -70.152 & 64.991 & 0.220 \nl
261094 & 22:10:50 & 3259 & 109.06 & 33.56 & 0.85 & 9.376 & 70.489 & 62.802 & 0.225 \nl
311094 & 11:21:58 & 3269 & 217.27 & -24.16 & 1.13 & 197.728 & -72.274 & 49.713 & 0.266 \nl
191194 & 19:30:50 & 3287 & 182.50 & 21.98 & 0.42 & 70.723 & 74.969 & 73.825 & 0.018 \nl
211194 & 17:24:57 & 3290 & 35.45 & -66.37 & 1.98 & 256.325 & -74.588 & 34.074 & 0.032 \nl
251194 & 22:04:30 & 3295 & 240.92 & -18.48 & 0.70 & 267.701 & -73.352 & 55.459 & 0.077 \nl
261194 & 12:22:01 & 3298 & 264.74 & -61.30 & 0.23 & 269.192 & -73.135 & 14.236 & 0.054 \nl
271194 & 14:30:17 & 3303 & 69.17 & 44.97 & 1.29 & 91.859 & 72.712 & 30.076 & 0.085 \nl
281194 & 18:41:26 & 3306 & 19.57 & -33.95 & 1.06 & 274.576 & -72.231 & 63.385 & 0.030 \nl
281294 & 07:39:41 & 3330 & 353.78 & -49.23 & 1.64 & 313.638 & -54.491 & 24.995 & 0.141 \nl
291294 & 22:50:58 & 3334 & 255.18 & 39.99 & 1.16 & 134.838 & 53.382 & 76.440 & 0.069 \nl
040195 & 07:18:39 & 3345 & 303.26 & -15.63 & 0.47 & 318.438 & -49.769 & 36.621 & 0.046 \nl
080195 & 02:04:20 & 3349 & 17.71 & -12.94 & 3.58 & 320.718 & -47.219 & 57.551 & 0.015 \nl
110195 & 04:47:43 & 3351 & 336.24 & -64.78 & 1.26 & 322.465 & -45.130 & 21.569 & 0.071 \nl
110195 & 12:18:30 & 3352 & 19.57 & -10.22 & 0.32 & 322.636 & -44.919 & 61.714 & 0.215 \nl
310195 & 11:35:35 & 3385 & 17.89 & 19.54 & 1.61 & 331.814 & -31.836 & 67.978 & 0.109 \nl
080295 & 02:10:24 & 3408 & 336.96 & 54.19 & 0.33 & 334.717 & -27.012 & 83.394 & 0.007 \nl
100295 & 02:20:21 & 3410 & 154.55 & -27.48 & 1.15 & 155.444 & 25.748 & 53.632 & 0.008 \nl
110295 & 02:24:57 & 3412 & 9.51 & 52.65 & 1.08 & 335.804 & -25.124 & 85.830 & 0.006 \nl
110295 & 20:07:48 & 3415 & 64.35 & -55.02 & 1.16 & 336.065 & -24.664 & 67.625 & 0.016 \nl
210295 & 20:45:05 & 3436 & 294.31 & -32.15 & 0.60 & 339.467 & -18.452 & 42.614 & 0.220 \nl
260295 & 14:48:56 & 3442 & 158.99 & -13.90 & 0.60 & 160.996 & 15.516 & 29.579 & 0.100 \nl
020395 & 23:52:54 & 3449 & 206.65 & 60.22 & 2.80 & 162.370 & 12.830 & 62.763 & 0.111 \nl
050395 & 15:05:04 & 3458 & 197.20 & -11.12 & 0.29 & 163.182 & 11.209 & 40.913 & 0.042 \nl
250395 & 07:10:44 & 3480 & 52.81 & 50.50 & 0.58 & 349.026 & 1.021 & 71.668 & 0.011 \nl
250395 & 17:36:31 & 3481 & 66.22 & -11.56 & 0.33 & 349.154 & 1.299 & 79.595 & 0.006 \nl
270395 & 07:19:07 & 3485 & 54.12 & 38.01 & 1.32 & 349.610 & 2.296 & 70.902 & 0.374 \nl
270395 & 23:47:51 & 3486 & 274.40 & 20.46 & 2.57 & 349.812 & 2.735 & 76.492 & 0.519 \nl
010495 & 11:50:42 & 3488 & 197.93 & 56.95 & 1.22 & 171.120 & -5.632 & 66.998 & 0.014 \nl
010495 & 20:43:54 & 3489 & 86.06 & 43.75 & 0.29 & 171.229 & -5.870 & 89.812 & 0.041 \nl
030495 & 13:19:47 & 3491 & 36.39 & 10.71 & 0.43 & 351.723 & 6.965 & 43.887 & 0.013 \nl
030495 & 23:33:46 & 3492 & 184.64 & 56.55 & 1.67 & 171.848 & -7.251 & 62.959 & 0.007 \nl
180495 & 23:16:35 & 3512 & 92.40 & 3.26 & 0.65 & 176.274 & -17.317 & 81.419 & 0.106 \nl
210495 & 12:29:49 & 3516 & 72.01 & -63.06 & 0.79 & 177.047 & -19.094 & 81.042 & 0.054 \nl
250495 & 00:15:19 & 3523 & 163.16 & -35.25 & 0.35 & 178.117 & -21.561 & 16.998 & 0.037 \nl
030595 & 18:36:10 & 3537 & 159.16 & -7.19 & 0.54 & 180.893 & -27.907 & 32.533 & 0.010 \nl
100595 & 07:47:59 & 3569 & 188.57 & -65.48 & 0.80 & 183.070 & -32.803 & 31.251 & 0.064 \nl
130595 & 22:41:43 & 3571 & 271.60 & -72.91 & 0.73 & 184.322 & -35.558 & 57.903 & 0.033 \nl
210595 & 06:59:17 & 3588 & 22.21 & -26.86 & 1.75 & 7.005 & 41.245 & 67.579 & 0.456 \nl
220595 & 23:41:23 & 3593 & 109.56 & 19.28 & 0.86 & 7.656 & 42.576 & 89.976 & 0.034 \nl
230595 & 05:39:29 & 3594 & 85.63 & 43.08 & 1.81 & 7.754 & 42.769 & 54.744 & 0.138 \nl
080695 & 22:34:32 & 3634 & 57.56 & -53.60 & 0.76 & 195.210 & -56.006 & 66.277 & 0.140 \nl
100695 & 01:54:09 & 3637 & 100.10 & -17.65 & 0.83 & 195.810 & -56.903 & 79.794 & 0.464 \nl
190695 & 05:28:39 & 3643 & 290.94 & -13.39 & 12.98 & 201.402 & -64.026 & 81.112 & 0.287 \nl
200695 & 00:44:33 & 3644 & 45.72 & 6.13 & 1.49 & 21.969 & 64.638 & 61.737 & 0.019 \nl
240695 & 23:21:24 & 3648 & 165.37 & 14.99 & 0.59 & 205.921 & -68.350 & 89.790 & 0.039 \nl
250695 & 04:03:21 & 3649 & 271.64 & -21.15 & 0.57 & 206.097 & -68.497 & 61.270 & 0.058 \nl
300695 & 21:10:37 & 3655 & 174.28 & 21.48 & 2.61 & 32.000 & 72.602 & 81.349 & 0.169 \nl
010795 & 03:32:38 & 3657 & 320.14 & -20.82 & 0.26 & 212.326 & -72.789 & 75.052 & 0.007 \nl
010795 & 06:35:37 & 3658 & 346.27 & 37.74 & 0.36 & 32.474 & 72.874 & 39.854 & 0.022 \nl
060795 & 11:52:15 & 3662 & 28.61 & 31.52 & 0.81 & 40.076 & 76.364 & 45.942 & 0.385 \nl
110795 & 03:49:49 & 3663 & 85.81 & -24.01 & 1.03 & 229.867 & -79.179 & 73.994 & 0.113 \nl
130795 & 08:56:16 & 3664 & 286.67 & 30.04 & 2.14 & 56.024 & 80.361 & 63.327 & 0.303 \nl
260795 & 14:19:38 & 3709 & 12.54 & -40.72 & 1.81 & 300.441 & -83.254 & 47.559 & 0.021 \nl
040895 & 01:58:54 & 3734 & 287.45 & 56.26 & 0.32 & 154.235 & 80.709 & 40.979 & 0.010 \nl
050895 & 03:44:14 & 3736 & 79.86 & -39.38 & 1.75 & 337.035 & -80.272 & 48.132 & 0.020 \nl
090895 & 23:52:25 & 3750 & 51.09 & 14.06 & 0.63 & 166.942 & 78.209 & 79.289 & 0.060 \nl
180895 & 01:23:22 & 3765 & 278.65 & 33.87 & 0.35 & 177.565 & 74.696 & 61.035 & 0.015 \nl
040995 & 14:38:08 & 3776 & 258.86 & 47.48 & 0.47 & 190.635 & 67.770 & 40.016 & 0.071 \nl
090995 & 23:44:12 & 3788 & 216.36 & -0.11 & 0.37 & 193.474 & 65.934 & 65.317 & 0.037 \nl
021095 & 21:33:06 & 3843 & 46.05 & -1.88 & 0.63 & 23.193 & -59.716 & 59.919 & 0.146 \nl
101095 & 21:12:23 & 3857 & 126.55 & 39.35 & 3.18 & 206.055 & 58.154 & 49.436 & 0.274 \nl
141095 & 12:06:04 & 3866 & 167.85 & -20.64 & 1.15 & 207.291 & 57.546 & 84.406 & 0.031 \nl
161095 & 00:41:21 & 3870 & 35.82 & -24.36 & 0.35 & 27.801 & -57.306 & 33.071 & 0.010 \nl
161095 & 19:53:28 & 3871 & 118.99 & -49.09 & 2.27 & 28.067 & -57.187 & 49.121 & 0.218 \nl
191095 & 13:24:04 & 3875 & 161.51 & -87.81 & 1.80 & 28.962 & -56.801 & 33.350 & 0.602 \nl
021195 & 05:52:55 & 3891 & 138.57 & 46.99 & 0.53 & 213.224 & 55.399 & 43.599 & 0.029 \nl
041195 & 01:37:42 & 3893 & 40.39 & 27.26 & 0.42 & 33.757 & -55.279 & 80.709 & 0.053 \nl
071195 & 17:32:37 & 3900 & 11.99 & 29.34 & 5.26 & 214.811 & 55.089 & 87.309 & 0.101 \nl
121195 & 18:27:59 & 3905 & 132.52 & 49.08 & 2.94 & 216.218 & 54.931 & 52.665 & 0.025 \nl
131195 & 23:25:55 & 3906 & 355.11 & -44.76 & 1.12 & 36.540 & -54.909 & 30.208 & 0.120 \nl
171195 & 02:48:38 & 3909 & 285.79 & 37.05 & 1.96 & 217.381 & 54.892 & 50.236 & 0.052 \nl
191195 & 08:17:12 & 3912 & 306.46 & -40.71 & 0.42 & 37.957 & -54.907 & 59.571 & 0.047 \nl
241195 & 05:54:18 & 3918 & 73.32 & 51.68 & 1.15 & 219.187 & 55.028 & 69.352 & 0.215 \nl
021295 & 10:51:40 & 3929 & 328.06 & -35.76 & 0.62 & 41.082 & -55.493 & 51.555 & 0.051 \nl
031295 & 01:06:27 & 3930 & 279.84 & -19.25 & 0.19 & 221.213 & 55.539 & 89.971 & 0.016 \nl
081295 & 05:26:56 & 3936 & 353.00 & 70.15 & 0.60 & 222.287 & 56.016 & 48.493 & 0.016 \nl
081295 & 11:47:23 & 3937 & 57.66 & -28.15 & 4.59 & 42.331 & -56.043 & 25.744 & 0.112 \nl
081295 & 23:24:13 & 3938 & 72.65 & -68.48 & 1.98 & 42.428 & -56.095 & 19.484 & 0.489 \nl
131295 & 04:57:32 & 3954 & 283.98 & -6.82 & 0.60 & 223.214 & 56.597 & 82.982 & 0.024 \nl
191295 & 16:51:19 & 4039 & 240.79 & 60.51 & 0.66 & 224.261 & 57.531 & 11.292 & 2.297 \nl
201295 & 08:51:21 & 4048 & 104.53 & 23.20 & 0.48 & 224.353 & 57.639 & 85.179 & 0.031 \nl
271295 & 22:00:00 & 4146 & 68.61 & 1.59 & 1.85 & 45.226 & -58.994 & 64.010 & 0.136 \nl
281295 & 15:47:26 & 4157 & 93.13 & 28.22 & 0.81 & 225.298 & 59.142 & 85.093 & 0.045 \nl
110196 & 12:55:23 & 4312 & 138.07 & 48.17 & 0.53 & 225.658 & 62.292 & 46.594 & 0.010 \nl
140196 & 12:15:04 & 4368 & 216.19 & -29.73 & 0.09 & 225.469 & 63.054 & 89.387 & 0.041 \nl
190196 & 09:37:55 & 4462 & 86.33 & 54.65 & 0.77 & 224.911 & 64.359 & 55.420 & 0.050 \nl
240196 & 00:56:25 & 4556 & 51.04 & 54.05 & 0.30 & 224.023 & 65.642 & 60.001 & 0.012 \nl
010296 & 21:48:43 & 4701 & 358.05 & 14.46 & 0.31 & 41.148 & -68.141 & 89.132 & 0.014 \nl
020296 & 14:05:51 & 4710 & 215.04 & -64.63 & 1.54 & 40.852 & -68.331 & 47.716 & 0.116 \nl
060296 & 04:58:17 & 4757 & 195.53 & -9.07 & 0.40 & 219.081 & 69.332 & 81.557 & 0.033 \nl
100296 & 09:46:37 & 4814 & 232.30 & 77.01 & 1.63 & 216.549 & 70.451 & 6.653 & 0.268 \nl
160296 & 16:03:16 & 4898 & 355.45 & 16.75 & 0.52 & 211.691 & 71.973 & 82.540 & 0.020 \nl
290296 & 23:56:04 & 5080 & 358.22 & 12.40 & 2.55 & 16.953 & -74.180 & 78.054 & 0.040 \nl
160396 & 18:34:44 & 5255 & 223.00 & -70.17 & 1.40 & 355.865 & -74.031 & 31.819 & 0.250 \nl
190396 & 14:26:33 & 5277 & 94.84 & -47.72 & 2.01 & 352.443 & -73.674 & 48.164 & 0.050 \nl
210396 & 21:11:20 & 5299 & 7.56 & 67.05 & 0.25 & 169.850 & 73.326 & 41.277 & 0.007 \nl
220396 & 05:27:21 & 5304 & 96.98 & -54.61 & 0.26 & 349.471 & -73.264 & 44.180 & 0.025 \nl
310396 & 05:53:26 & 5389 & 111.86 & -20.39 & 0.85 & 341.046 & -71.409 & 80.424 & 0.047 \nl
080496 & 22:12:27 & 5413 & 85.04 & 1.88 & 1.10 & 155.435 & 69.173 & 83.270 & 0.173 \nl
090496 & 19:56:17 & 5416 & 15.52 & 42.42 & 1.37 & 154.975 & 68.924 & 63.480 & 0.077 \nl
090496 & 21:25:44 & 5417 & 42.54 & 41.92 & 1.34 & 154.944 & 68.907 & 58.168 & 0.016 \nl
150496 & 21:41:28 & 5427 & 2.64 & 15.12 & 2.06 & 332.405 & -67.200 & 86.464 & 0.117 \nl
160496 & 04:08:59 & 5428 & 66.85 & 73.60 & 2.06 & 152.315 & 67.127 & 25.706 & 0.199 \nl
180496 & 01:41:44 & 5433 & 170.50 & 0.42 & 1.46 & 151.681 & 66.572 & 69.690 & 0.042 \nl
180496 & 18:33:15 & 5436 & 110.32 & -16.15 & 0.70 & 151.469 & 66.367 & 88.460 & 0.008 \nl
220496 & 16:39:35 & 5443 & 134.80 & 27.20 & 1.50 & 150.446 & 65.217 & 40.657 & 0.100 \nl
250496 & 23:59:50 & 5447 & 9.80 & 56.51 & 1.46 & 149.780 & 64.243 & 55.057 & 0.021 \nl
280496 & 13:12:32 & 5450 & 304.24 & 35.13 & 1.01 & 149.374 & 63.491 & 77.240 & 0.065 \nl
300496 & 13:51:26 & 5451 & 11.80 & -51.49 & 1.11 & 329.116 & -62.893 & 28.853 & 0.137 \nl
160596 & 23:24:08 & 5464 & 62.28 & -18.07 & 1.82 & 328.560 & -58.205 & 77.589 & 0.150 \nl
230596 & 16:32:01 & 5470 & 46.75 & -63.32 & 1.85 & 328.920 & -56.373 & 38.872 & 0.035 \nl
240596 & 09:05:26 & 5472 & 160.82 & 2.60 & 2.83 & 148.969 & 56.192 & 54.514 & 0.349 \nl
240596 & 20:09:09 & 5473 & 358.52 & -24.77 & 0.72 & 329.006 & -56.066 & 38.585 & 0.032 \nl
280596 & 02:14:22 & 5476 & 291.86 & 22.03 & 1.49 & 329.287 & -55.209 & 84.466 & 0.025 \nl
290596 & 12:07:45 & 5477 & 14.40 & 48.89 & 0.36 & 149.423 & 54.842 & 71.176 & 0.004 \nl
310596 & 01:54:35 & 5479 & 34.21 & -28.50 & 1.32 & 329.587 & -54.434 & 53.620 & 0.031 \nl
050696 & 08:09:52 & 5486 & 154.26 & -1.98 & 0.23 & 150.195 & 53.106 & 52.945 & 0.034 \nl
070696 & 21:41:15 & 5489 & 48.43 & 75.10 & 0.36 & 150.528 & 52.471 & 41.496 & 0.015 \nl
100696 & 21:23:02 & 5492 & 328.24 & -29.35 & 2.63 & 330.940 & -51.742 & 23.798 & 0.667 \nl
230696 & 01:18:48 & 5512 & 180.78 & 7.59 & 0.74 & 152.847 & 48.923 & 46.048 & 0.046 \nl
230696 & 09:11:43 & 5514 & 163.70 & -59.73 & 5.98 & 332.903 & -48.848 & 74.746 & 0.026 \nl
240696 & 03:26:13 & 5517 & 98.96 & 32.05 & 1.83 & 153.032 & 48.680 & 43.631 & 0.221 \nl
250696 & 04:42:26 & 5518 & 110.59 & -8.21 & 0.89 & 153.213 & 48.448 & 68.138 & 0.109 \nl
280696 & 05:20:22 & 5523 & 326.98 & 48.28 & 1.22 & 153.745 & 47.787 & 83.868 & 0.030 \nl
020796 & 15:11:26 & 5525 & 36.78 & -15.53 & 0.47 & 334.542 & -46.848 & 58.984 & 0.110 \nl
030796 & 13:42:53 & 5526 & 4.62 & -7.75 & 0.96 & 334.713 & -46.652 & 46.702 & 0.136 \nl
070796 & 10:16:40 & 5530 & 320.95 & 82.51 & 1.07 & 155.435 & 45.862 & 50.112 & 0.027 \nl
070796 & 12:24:53 & 5531 & 189.27 & -1.78 & 1.23 & 155.451 & 45.843 & 53.860 & 0.121 \nl
080796 & 20:09:32 & 5534 & 307.03 & 9.80 & 0.51 & 335.703 & -45.574 & 61.292 & 0.018 \nl
150796 & 00:48:26 & 5539 & 3.43 & 35.50 & 1.69 & 336.899 & -44.355 & 85.016 & 0.096 \nl
220796 & 07:05:06 & 5548 & 29.64 & 31.67 & 0.91 & 338.339 & -42.989 & 88.901 & 0.012 \nl
310796 & 05:46:01 & 5557 & 43.93 & 12.51 & 0.89 & 340.144 & -41.403 & 83.839 & 0.012 \nl
030896 & 18:45:22 & 5561 & 338.02 & -50.13 & 1.93 & 340.865 & -40.801 & 12.019 & 0.014 \nl
040896 & 23:28:55 & 5563 & 218.67 & 68.20 & 0.39 & 161.109 & 40.606 & 41.491 & 0.004 \nl
070896 & 19:49:26 & 5567 & 157.53 & 32.13 & 0.21 & 161.690 & 40.139 & 10.156 & 0.025 \nl
080896 & 16:42:03 & 5568 & 85.25 & 37.02 & 0.44 & 161.869 & 39.998 & 57.917 & 0.009 \nl
150896 & 10:00:27 & 5575 & 265.33 & -58.01 & 0.98 & 343.241 & -38.943 & 56.214 & 0.008 \nl
\enddata
\end{deluxetable}

\end{document}